\documentstyle[12pt]{article}
\author{\\ \normalsize
E.~L.~Afraimovich, O.~I.~Berngardt, O.~S.~Lesyuta, A.~P.~Potekhin
and B.~G.~Shpynev \\
\vspace*{0.3cm}
\\
\sl Institute of Solar-Terrestrial Physics, Russian Academy of Sciences,\\
\sl 664033~~Post Box~4026, Irkutsk, Russia,
e-mail:afra@iszf.irk.ru}
\title{\vspace{2.3cm} \Large {\bf
A Case Study of The Mid-Latitude GPS Performance at Nighttime
During The Magnetic Storm of July 15, 2000}}
\date{}
\begin{document}
\maketitle
\begin{abstract}
Using the geomagnetic storm of July 15, 2000 as an example, we
investigated the dependence of GPS navigation system performance
on the nightside at mid-latitudes on the level of geomagnetic
disturbance. The investigation was based on the data from the
global GPS system available through the Internet. It was shown
that the number of GPS phase slips increases with the increasing
level of disturbance and that there is a good correlation between
the rate of $Dst$-variation and the frequency of slips. It was
further shown that the relative frequency of slips has also a
clearly pronounced aspect dependence. Phase slips of the GPS
signal can be caused by the scattering from small-scale
irregularities of the ionospheric E-layer. Phase slip
characteristics are indicative of Farley-Buneman instabilities as
a plausible physical mechanism that is responsible for the
formation of geomagnetic field-aligned irregularities. Using
simultaneous measurements of backscatter signal characteristics
from the Irkutsk incoherent scatter radar and existing models for
such irregularities, we estimated the order of magnitude of the
expected phase fluctuations of the GPS signal at a few degrees.
\end{abstract}

\section{Introduction}

The satellite navigation GPS system has become a powerful factor
of scientific and technological progress worldwide, and enjoys
wide use in a great variety of human activity. In this connection,
much attention is given to continuous perfection of the GPS system
and to the widening of the scope of its application for solving
the navigation problems themselves, as well as for developing
higher-precision systems for time and accuracy determinations.
Even greater capabilities are expected in the near future through
the combined use of the GPS with a similar Russian system
(GLONASS). The operation of the GPS is considerably affected by
characteristics of the medium lying along the path of signal
propagation. Because of this it is important to analyze the GPS
performance during strong geomagnetic disturbances that change the
characteristics of one of the regions having the greatest
influence on GPS signals, the ionosphere. Failures in the GPS
operation in the polar and equatorial regions are frequently
discussed (in, e.g., Skone at al., 2000). However, the presence of
failures at mid-latitudes is scantily discussed in the literature
(Afraimovich et al., 2000), in spite of reasonably complete
spatial coverage and the extensive practical applications of GPS
receivers just in the mid-latitude region.

This paper considers the dependence of the time and space
distribution of phase slips (PS) of the mid-latitude GPS receivers
on the nightside of the Earth based on using data from the Irkutsk
incoherent scatter (IS) radar and time-coincident two-frequency
phase measurements from the mid-latitude GPS receivers in the
Earth's northern hemisphere obtained during the geomagnetic
disturbance on July 15, 2000 (17-24 UT (July 15, 2000), 00-07 of
Irkutsk LT (July 16, 2000), with maximum values of Kp=9, and
$Dst$=-325 nT) (Afraimovich et al., 2000).

\section{Experimental layout}

A preprocessing of the GPS data provided estimates of the
normalized (to a total number of observations) relative density of
phase slips (RDPS). Ascertaining the reason behind the increase in
the slip density was also greatly facilitated by the intensity
estimates of total electron content (TEC) variations  for the same
stations and time intervals determined by a standard technique of
two-frequency measurements. The acquired data were used to derive
the mean power spectrum $<S^{2}(F)>$ of TEC variations - the
result of an averaging of the energy spectrum of TEC variation for
each path of the subionospheric point for all such paths.
Throughout the text below, we shall use the TEC unit TECU equal to
$10^{16}$ $m^{-2}$ which is generally accepted in the literature.

We have also calculated the dependence of the RDPS as a function
of line-of-sight (LOS) azimuth and elevation from the GPS receiver
to the satellite. The azimuth was measured from the northward
direction, and the elevation from the horizontal (is approximately
equal to the cotangent of the ratio of the height of the
subionospheric point to the range to it). For comparison, the data
for the magnetically quiet day of July 29, 1999 were processed,
during which the geomagnetic disturbance $Dst$ index was on the
order of -10 nT).

The Irkutsk IS radar is located at 120 km north of Irkutsk
($53^\circ$N, $104^\circ$E); it is a monostatic facility for
probing the mid-latitude ionosphere using the radio wave
backscattering method at 154-162 MHz frequencies (Potekhin et al.,
1999). The radar includes the linearly polarized receive-transmit
antenna for making, based on the IS method (Evans, 1969), absolute
measurements of the electron density from Faraday fadings of the
received signal, as well as electron and ion temperatures and the
plasma drift velocity in the ionospheric F-region (height range
170-750 km). During geomagnetic disturbances the radar can be used
(Potekhin et al., 1999) to investigate the ionospheric E-layer
(height range 100-120 km) using data on backscatter (BS) from
small-scale irregularities extended along the Earth's magnetic
field (Haldoupis et al., 1997).

The antenna beam of the radar is such that when investigating the
ionospheric F-region by the IS method, the scattered signal is
received by the main lobe of the antenna beam. When investigating
the E-layer by the BS method, the signal is received  by the lower
sidelobes of the beam, with a power attenuation on the order of
$10^{-11}$ with respect to the main lobe. The experiment under
discussion involved measurements of the IS signal power, the BS
signal power, and of the plasma drift velocity.

\section{Description and discussion of the GPS data}

The data presented in Fig.~1 clearly show a high correlation
between the time derivative of the geomagnetic disturbance
$Dst$-index and the number of PS. The number of phase slips
exceeds one percent in the disturbance maximum, and the position
of the maximum is closely time-coincident with maximum absolute
values of $dDst/dt$. The data in Fig. 1a also intimate that on the
magnetically quiet day of July 29, 1999, the number of PS
decreases by several orders of magnitude (Fig.~1a, gray line with
stars), compared with the number of PS during disturbed periods
(Fig.~1a, black line). This suggests that the phase slips of the
GPS receivers are associated with an increase of the level of
geomagnetic disturbance and can be caused by phase distortions of
GPS signals as they propagate through a disturbed region of the
ionosphere that is abundant in irregularities of different scales.

Fig.~2 presents the spectral power $<S^{2}(F)>$ of TEC variations
(Fig.~2a), and the dependence of the relative density of PS P(N,E)
on the position of the subionospheric point with respect to the
GPS station's location (Fig.~2b). It is evident from Fig.~2 that
during the geomagnetic storm of July 15, 2000 (20-22 UT) the TEC
variation power increases by 1-2 orders of magnitude, both in the
low- and high-frequency parts of the spectrum. This suggests that
the slips can be caused by phase distortions of GPS signals under
the influence of small-scale irregularities (smaller than or on
the order of the radius of the Fresnel zone), the amplitude of
which increases with increasing geomagnetic disturbance level.
Fig. 2b clearly shows a strong aspect dependence of the RDPS on
the position of the subionospheric point (the direction of the
beam to the satellite) - PS occur most commonly in the northward
direction, at low elevations of the "satellite-receiver" beam
($\alpha$$<$300).

\section{Description and discussion of the Irkutsk IS radar data}

One of the best-known types of small-scale irregularities, the
scattering of which is characterized by a strong aspect
dependence, are the geomagnetic field-aligned E-layer
irregularities (Buneman, 1963; Farley, 1963).

Fig.~3 presents the data from the Irkutsk IS radar: the dependence
of the scattered signal power and ionospheric plasma drift
velocity as functions of time and radar range R. The measurements
of the backscattered signal power were made with respect to the IS
signal level corresponding in the figure to the 10 dB level. The
drift velocity measurements were made from the frequency Doppler
shift of the scattered signal (Haldoupis, 1997).

An analysis for a similar storm of September 25, 1998 showed that
the strongest scattering occurring within the ranges of 550-1100
km corresponds to the scattering from the geomagnetic
field-aligned ionospheric E-layer irregularities (Potekhin et al.,
1999). Such irregularities are usually observed in the
mid-latitude ionosphere during strong geomagnetic disturbances
(Foster and Tetenbaum, 1991; Haldoupis, 1997). It is evident from
Fig. 3b that between 22 and 24 UT the drift velocity exceeded the
ion-sound velocity of 250 m/s; for that reason, the necessary
conditions (Buneman, 1963; Farley, 1963) for the generation of
such irregularities were satisfied. It was shown (Potekhin et al.,
1999) that the signal that is scattered from E-layer
irregularities is received by the lower sidelobes of the antenna
beam in the direction northward of the radar; for that reason, the
power of this BS signal is attenuated by 11 orders of magnitude
due to the experimental geometry. The IS signal is received by the
main beam lobe; therefore, it does not undergo such an additional
attenuation.

A comparison of Fig.~3a and Fig.~1 reveals that the occurrence of
an anomalously powerful scattering from geomagnetic field-aligned
E-layer irregularities, maximum values of the derivative $dDst/dt$
and of the relative density of phase slips of GPS receivers are
time-coincident.

The relation of the scattering cross-section of irregularities to
the scattered signal power is defined by the radar equation
(Isimaru, 1978) (within a factor that is unimportant for a further
consideration):

\begin{equation}
\label{SliHR-eq-01} P_{rc} (\vec R) = const \cdot P_{tr}
\frac{{G^2 (\vec R)}}{{R^2 }}\sigma (\vec R)V
\end{equation}

where $P_{rc}$ and $P_{tr}$ stand for the power of the scattered
and sounded signals, respectively; $G$ is the antenna power gain
in the direction of the sounding volume; $\sigma$  is the
scattering section per unit volume; and $V$ is the sounded volume.

The radar equation (1) can be used  to estimate the relative
scattering cross-section of irregularities, from which the
scattering occurs (taking into consideration that the values of
$R$, $V$ and $P_{tr}$ are of the same order of magnitude for the
two different scattering mechanisms):

\begin{equation}
\sigma _{BS} \sim \frac{{P_{rc,BS} }}{{P_{rc,IS} }}\frac{{G_{IS}
^2 }}{{G_{BS} ^2 }}\sigma _{IS}.
\end{equation}

The power of the backscattered signal $P_{rc,BS}$ exceeds that of
the incoherently scattered signal $P_{rc,IS}$ by three orders of
magnitude ($P_{BS}/P_{IS}$=$10^3$, Fig.~3a). The antenna gain in
the two cases under consideration are related by the relation
$G_{IS} ^{2}/G_{BS} ^{2}$=$10^{11}$. The scattering section of the
IS signal is known (Evans, 1969): $\sigma _{IS} \sim 10^{-19}$
$m^{-1}$ . Therefore, from (2) one can estimate the scattering
section of geomagnetic field-aligned ionospheric E-layer
irregularities at the mean frequency of 154 MHz of the Irkutsk IS
radar.

\begin{equation}
\sigma _{BS} \sim 10^{-5} m^{-1}.
\end{equation}

\section{The possible cause of phase slips of GPS receivers}

One conceivable reason for PS of GPS receivers on the nightside
can be the phase distortions of the GPS signal caused by the
scattering from small-scale irregularities of ionospheric plasma.

The distribution $P(\gamma)$ of the relative density of P,
constructed as a function of the angle $\gamma$ between the
propagation direction of the GPS signal and the geomagnetic field,
has a pronounce tendency for an increase in relative density of PS
as $\gamma$ approaches $90^\circ$ (Fig.~4a). This result is also
in good agreement with the feature pointed out in Section 3, the
increase in the slip density in the case of the northward directed
LOS to the satellite at low elevations $\alpha$.

We carried out a numerical simulation of the scattering from
geomagnetic field-aligned irregularities using the international
reference model of the magnetic field (IGRF) and the well-known
model of irregularities (Uspensky and Starkov, 1987) that has an
approximate form:

\begin{equation}
\sigma _{BS} (\vec R) = \sigma _0 e^{ - ((h(\vec R) - 110)/10)^2 }
10^{ - |\gamma (\vec R) - 90^\circ |/10}
\end{equation}

where $h(\vec R)$ and $\gamma (\vec R)$ are the altitude of the
point $\vec R$ above ground level (in km) and the angle with the
magnetic field (in degrees), respectively. To obtain $P(\gamma)$,
the expression (1) was integrated with respect to all directions
of the vector $\vec R$ in view of (4) and the directional pattern
of the Irkutsk IS radar $G(\vec R)$. The simulation showed that
the strongest scattering occurs in the region close to the region
of the experimentally observed maximum of the RDPS (Fig.~3b,~4b).
This suggests that it is the scattering from the geomagnetic
field-aligned E-layer irregularities which causes phase slips of
GPS receivers.

Let us now estimate the level of phase distortions in terms of the
proposed mechanism. By approximating the scattering section of the
oriented irregularities by the experimental law $\ln (\sigma _{BS}
(k_1 )) = a_0  + a_1 k^{ - 2.25}$ obtained in (Moorcroft, 1987),
from (3) we can estimate the absolute scattering section per unit
volume at the frequency $f_1$ = 1575{.}42 MHz of GPS performance:

\begin{equation}
\sigma _{1575}  = \sigma _{BS} \left( {\frac{{150}}{{1575}}}
\right)^{2.25} \sim 10^{ - 8}  - 10^{ - 7} m^{ - 1}.
\end{equation}

In accordance with Rytov's method (Isimaru, 1978), phase
distortions of the signal in a medium with the irregularity
section $\sigma$ per unit volume and with the propagation path
length $L$ in this medium, are defined by the expression:

\begin{equation}
< \Delta \varphi ^2  >  = L\sigma /2.
\end{equation}

Taking into account the thickness of the E-layer $L\sim10^{4}$ m,
it is possible to estimate minimum phase distortions of the GPS
signal:

\begin{equation}
\Delta \varphi \sim 3\sqrt { < \Delta \varphi ^2  > } \sim 10^{ -
1} - 10^{ - 2} rad,
\end{equation}

taking into account the oblique propagation through the E-layer,
$\Delta \varphi$ will be still larger.

Thus, in accordance with the proposed model, phase distortions of
the signal can exceed a few degrees and lead, even in the case of
such a noise-stable system as the GPS, to phase slips with a
standard pretreatment of the GPS signal.

\section{Conclusion}

In this paper we have considered some of the space and time
characteristics of phase slips of mid-latitude GPS receivers on
the nightside of the Earth during the strong geomagnetic
disturbance of July 15, 2000.

It has been shown that the number of PS increases with the
increasing rate of $Dst$-variation with the time, and has a
clearly pronounced aspect dependence - the relative frequency of
phase slips is the largest in the northward direction at small
elevations. We have suggested a plausible mechanism that is
responsible for PS - phase distortions of the GPS signal
propagating through the ionosphere that are caused by the
scattering from the geomagnetic field-aligned E-layer
irregularities. Using the data from the Irkutsk IS radar we have
estimated the order of phase distortions of the GPS signal that
are caused by this mechanism and amounting to several degrees.

To reliably explain the night-time phase slips of GPS receivers in
terms of this mechanism alone requires a more detailed
verification of the functioning of GPS receivers produced by
different manufacturers, for such a level of phase distortions,
and a relevant numerical simulation of the algorithms involved.

\section*{Refereences}
{\hspace{-0.2cm}\makebox{
\parbox{13.5cm}{
\hspace{0.5cm}Afraimovich~E.~L., Lesyuta~O.~S., and Ushakov~I.~I.,
Magnetospheric disturbances, and the GPS operation, LANL e-print
archive, http://xxx.lanl.gov/abs/physics/0009027, 2000.}}}

\vspace{0.1cm}

Buneman~O., Excitation of field-aligned sound waves by electron
streams, Phys. Råv. Lett., V.10, 285--287, 1963.

Evans~J.~V., Theory and Practice of Ionosphere study by Thomson
Scatter Radar, Proc. IEEE, V. 57, 496--530, 1969.

Farley~D.~T., A plasma instability resulting in field-aligned
irregularities in the ionosphere, J. Geophys. Res., V.68,
6083--6097, 1963.

Foster~J.~C., and Tetenbaum~D., High resolution backscatter power
observations of 440-MHz E-region coherent echoes at
Millstone-Hill, J. Geophys. Res., V.96, 1251--1261, 1991.

Haldoupis~C., Farley~D.~T., and Schlegel~K., Type-1 echoes from
mid-latitude E-region ionosphere, Ann. Geophysicae, V.15,
908--917, 1997.

Ishimaru~A., Wave Propagation and Scattering in Random Media,
Academic Press, 1978.

Potekhin~A.~P., Berngardt~O.~I., Kurkin~V.~I., Shpynev~B.~G.,
Zherebtsov~G.~A., Foster~J.~C., and Rich~F.~J., Observation of the
Mid-Latitude Coherent Echoes During the September 25,1998 Storm
With the Irkutsk IS Radar, Abstracts of XXIV GA URSI, Toronto,
Canada, 441, 1999.

Moorcroft~D.~R., Estimates of absolute scattering coefficients of
radar aurora, J. Geophys. Res., V.92, 8723--8732, 1987.

Skone~S., de Jong~M., The impact of geomagnetic substorms on GPS
receiver performance, Earth Planets Space, V.52, 1067--1071, 2000.

Uspensky~M.~V., and Starkov~G.~V., Polar Auroras and Radio Wave
Scattering. Leningrad: Nauka, 1987 (in Russian).
\end{document}